\documentclass[aps,prd,showpacs,superscriptaddress,twocolumn]{revtex4}
\usepackage{epsfig}
\usepackage[english]{babel}
\usepackage{pgf}
\usepackage{amsfonts,amsmath,amssymb,enumerate,array,amssymb}
\usepackage[latin1]{inputenc}
\usepackage{bm}
\usepackage{fixmath}
\usepackage{amsbsy}
\usepackage{graphicx}
\usepackage{capt-of}
\usepackage{dcolumn}

\usepackage{layouts}
\usepackage{natbib}
\usepackage{abrev-jour-long}

\def\S{\mathcal{S}}

\newcommand{\mom}{\mathcal{P}}
\newcommand{\DF}{\mathcal{F}}

\newtheorem{theorem}{Proposition}
\newtheorem{lema}{Lemma}

\begin{document}

\title{Distribution functions for a family of
general-relativistic Hypervirial models in collisionless regime}

\author{Henrique Matheus Gauy}\email[]{henmgauy@df.ufscar}

\author{Javier Ramos-Caro}\email[]{javier@ufscar.br}
\affiliation{Departamento de F\'isica, Universidade
Federal de S\~{a}o Carlos, S\~ao Carlos, 13565-905 SP, Brazil}


\begin{abstract}
By considering the Einstein-Vlasov system for static spherically symmetric
distributions of matter, we show that configurations with constant anisotropy
parameter $\beta$ have, necessarily, a distribution function (DF)
 of the form $\DF=l^{-2\beta}\xi(\varepsilon)$, where $\varepsilon=E/m$ and
  $l=L/m$ are the relativistic energy and angular momentum per unit rest mass, respectively.
We exploit this result to obtain
DFs for the general relativistic extension of the Hypervirial family introduced by Nguyen and Lingam (2013),
which Newtonian potential is given by $\phi(r)=-\phi_o /[1+(r/a)^{n}]^{1/n}$ ($a$ and $\phi_o$ are positive free parameters, $n=1,2,...$).
Such DFs can be written in the form
$\DF_{n}=l^{n-2}\xi_{n}(\varepsilon)$.
For odd $n$, we find that $\xi_n$  is a polynomial of order $2n+1$ in $\varepsilon$, as in the case of the Hernquist model ($n=1$),
for which $\DF_1\propto l^{-1}\left(2\varepsilon-1\right)\left(\varepsilon-1\right)^2$.
For even $n$, we can write $\xi_n$ in terms of incomplete beta functions (Plummer model, $n=2$, is an example).
Since we demand that
 $\DF\geq 0$ throughout the phase space, the particular form of each $\xi_n$ leads to restrictions for the values of $\phi_o$. For example, for the
Hernquist model we find that $0\leq \phi_o \leq2/3$, i.e. an upper bounding value less than the one obtained for Nguyen and Lingam ($0\leq \phi_o \leq1$), based on energy conditions.

%
%

\end{abstract}

\date{\today}

\maketitle

\section{Introduction}

Globular clusters, galactic bulges and dark matter haloes have been usually modeled as many-particle systems endowed by spherical symmetry.
Although the Newtonian theory of gravitation is usually chosen as one of the paradigms of galactic dynamics,
the idea of formulating these models in the general relativistic realm has been gaining interest in recent decades
\cite{0264-9381-10-6-003,lemosLetelier1994PRD,PhysRevD.62.064025,0264-9381-17-7-303,0264-9381-19-14-321,zacekSemerak2002CzJP,
vogtLetelier2005MNRAS,doi:10.1142/S0217751X0703666X,0264-9381-25-4-045011,loraclavijo-ospinahenao-pedraza2010PRD,
PhysRevD.86.043008,Carrick2012,1475-7516-2012-09-031,1742-6596-545-1-012006}
becoming one of the topical problems in stelar dynamics and relativistic astrophysics.

If one adopts a statistical standpoint to analyze such self-gravitating configurations,
it is advisable to perform the description by considering the Einstein-Vlasov system, in order to provide, in a self-consistent fashion,
the metric, the energy-momentum tensor and the distribution function (DF). In the context of galactic dynamics, usually based on Newtonian gravity,
these theoretical constructions are called as dynamical models: the set composed by DF, potential and density (see
\cite{binneyMcmillan2011MNRAS,binney2010MNRAS} for example).
 In this paper, adopting the
general relativistic paradigm, we also shall call the solutions of the Einstein-Vlasov system as dynamical models.

On one hand, the DF or probability density function, can be considered as a concept involving
all the relevant physical information about the system. Once the DF is known we can have access
to astrophysical observables as, for example, the projected density and the light-of-sight velocity, provided by photometric and kinematic measurements.
 On the other hand, the DF is a dynamical entity governed by a kinetic equation which determines the statistical evolution of the configuration.
For systems in collisionless regime it obeys the Vlasov equation, sometimes called as collisionless Boltzmann equation.
In the case of many-particle self-gravitating systems, the term ``collisionless'' is devoted
to situations where the gravitational encounters are not significant in the evolution. Important examples are galaxies and clusters of galaxies,
whose life time is lesser than the corresponding relaxation time. But for smaller systems as  stellar clusters, galactic bulges and haloes,
encounters might play a significant role in the evolution and the DF is said to obey the Fokker-Planck equation, which contains a
collision term characterized by the so-called \emph{diffusion coefficients}. Usually, they are computed by taking into account
an equilibrium DF that is solution of the Vlasov equation.

In other words, the task of describing the evolution of globular clusters in collision regime, starts with the knowledge of
the corresponding stationary DF in collisionless regime. Such a DF must determine, in a self-consistent manner, the associated energy-momentum
and metric tensors under equilibrium conditions.
In this line we will focus the principal subject of the present
 paper: providing adequate DFs, solutions of Einstein-Vlasov equations, for certain self-gravitating
spherically symmetric configurations of astrophysical interest in general relativity. For such purpose,
the well known \emph{$\rho$ to $f$ approach} of Newtonian gravity
\cite{1911MNRAS..71..460P,1916MNRAS..76..572E,1962MNRAS.123..447L,1983MNRAS.202..995J,binneytremaineGD,hernquist1990ApJ,pedrazaRamoscaroGonzalez2008MNRAS},
 which obtains the DF
starting from the potential-density pair, by inversion, can also be used in the
General Relativity realm. Here, we will show that for certain spherical distributions this procedure can be
performed analytically.

A wide variety of astrophysical configurations can be represented as
spherical systems with pressure anisotropy (the so-called anisotropic models), as confirmed by a
 number of authors in the last three decades
\cite{1974ApJ...188..657B,PhysRevD.26.1262,1992IJTP...31..545S,1994CQGra..11.2553C,doi:10.1063/1.532002,1997PhR...286...53H,HERRERA1998113,0264-9381-15-11-022,PhysRevD.62.104002,Dev:2000gt,doi:10.1139/p03-124,Mak393,PhysRevD.69.084026,Maharaj2006,MMA:MMA665,PhysRevD.77.027502,Sharma2012,PhysRevD.88.027301,PhysRevD.87.087303,doi:10.1093/mnras/stt773,PhysRevD.88.064020,Nguyen}. They are characterized by
an anisotropy parameter $\beta$ measuring the quotient between the radial pressure $P_r$ and the tangential (or azimuthal)
pressure $P_{\theta}$. In particular, for $\beta$ constant (i.e. independent of the radial coordinate $r$), it can be
proven that the DF is proportional to $L^{-2\beta}$ (see section \ref{sec:anisotropic}), as in the case of the hypervirial models
\cite{Nguyen}, for which $\beta=(2-n)/2$, with $n=1,2,...$, admitting some cases of interest.
For $n=1$ (the Hernquist model), since $\lim _{L\rightarrow 0}\DF=\infty$, radial orbits are
much more abundant that closed orbits and we expect most of the matter distribution to be located in the
inner region of the system. For $n>2$, the situation is the opposite: the DF increases with $L$,
leading to configurations with an overabundance of closed orbits and we do not expect a large mass concentration
near the center. The case $n=2$ (Plummer model) is the only isotropic model of this family, where the mass distribution tends to be homogeneous.
These features, along with the interesting property of satisfy  the virial theorem locally, makes the hipervirial
family a set of models appropriate to represent galaxies and dark matter halos, from both a Newtonian \cite{2005MNRAS.360..492E} and
relativistic \cite{PhysRevD.88.064020,Nguyen} point of view.

Apart from the characteristics mentioned above, the relativistic hypervirial models introduced
by \citeauthor{Nguyen} \cite{Nguyen} have the remarkable property of having the  same constant anisotropy parameter as their Newtonian counterparts.
Here we will exploit this fact to derive analytical expressions for the associated general-relativistic DFs determining the energy-momentum tensor
 and other basic settings making such models physically realizable configurations.
 In particular, it is worth mentioning that the requirement that the DFs be positive
leads to diminish the upper bounds of the free parameters (see section \ref{sec:phio}),
compared with the ones obtained from energy conditions \cite{Nguyen}.
In this sense, the requirement that the DFs be positive can be interpreted as a statement more fundamental than
the imposition of energy conditions  (an interesting analysis can also be found in \cite{Andreasson2011}).

The paper is organized as follows: In Sec. \ref{sec:general} we comment some general features of the relativistic extension
of Hernquist solution, focusing on the requirements that must hold to obtain physically realizable configurations, from the perspective
of energy conditions. We will show that they impose an upper bound of $4/3$ for the positive free parameter $\phi_o$. However this
upper limit decreases to $2/3$ with the knowledge of the DF (Sec. \ref{sec:DF-hernquist}).
In Sec. \ref{sec:selfgraveqs} we present a derivation of the self-gravitation equations (i.e. the Einstein-Vlasov  system) for
static, spherically symmetric distributions, in order to set the basis for the derivation of distribution functions, which
is performed in Secs. \ref{sec:DF-hernquist} (for the Hernquist solution) and \ref{sec:DF-hipervirial} (for the Hypervirial family).

Finally, some words on notation. Throughout the  paper we use natural units, $c=1$, where $c$ is the speed of light.
Greek indices $\mu,\nu$ run from $0$ to $3$. When using isotropic coordinates $(t,r,\theta,\psi)$ we
introduce the following associations for indices: $0\rightarrow t$, $1\rightarrow r$, $2\rightarrow \theta$ and
$3\rightarrow \psi$. Thus the symbol $T^{rr}$ will denote $T^{11}$, as well as $P_0$ equals to $P_t$, for example.

%

\section{A General-relativistic version for the Hernquist model}\label{sec:general}

\begin{figure*}[ht]
 \includegraphics[scale=0.55]{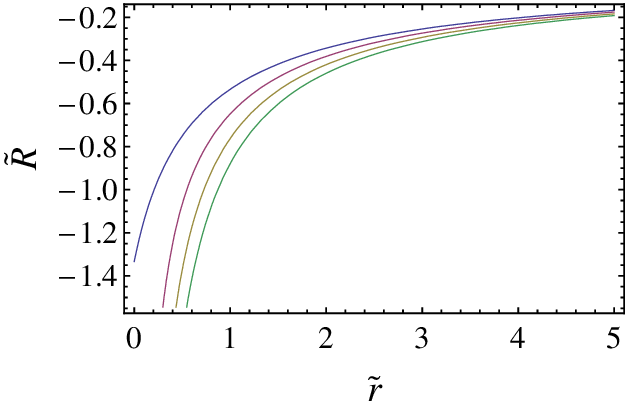}
 \includegraphics[scale=0.55]{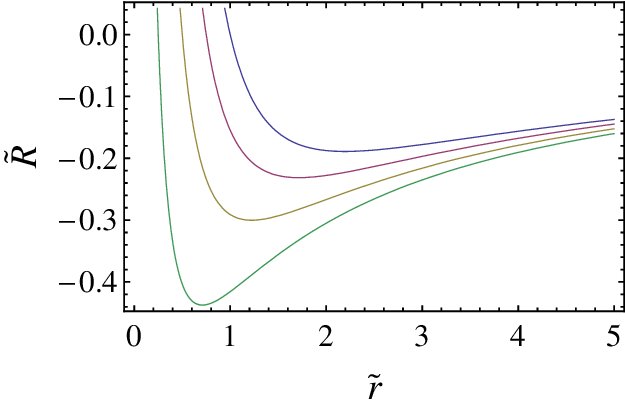}
 \includegraphics[scale=0.55]{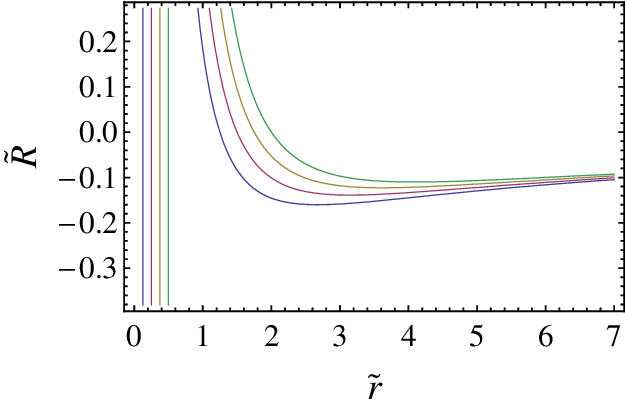}
  \caption{We show Ricci scalar for different values of parameter $\phi_o$. In particular we plot $\tilde{R}=(a^{3}/4\phi_o)R$ as
a function of $\tilde{r}=r/a$. For $0<\phi_o\leq 1$, we have $\tilde{R}<0$ (left panel). In the half panel we show $R$ for
 $1<\phi_o\leq 2$, which is positive only near the singularity $r=0$. For $\phi_o>2$ we have two singularities and also $R$ is negative in a prominent
 region of its domain.
  }
  \label{fig:Ricci}
\end{figure*}

The general static isotropic metric, in isotropic coordinates $(t,r,\theta,\psi)$,  can be written as \cite{Weinberg}
\begin{equation}\label{isotropicMetric}
\mathrm{d}s^2=-A(r)\mathrm{d}t^2 +B(r)\left(\mathrm{d}r^2+r^2\mathrm{d}\theta^2+r^2 \sin^2\theta\mathrm{d}\psi^2\right).
\end{equation}
Also, it can be expressed as a generalized version of the Schwarzschild metric, by defining
\begin{equation}\label{isotropicMetric2}
A(r)=\left[\frac{1-f(r)}{1+f(r)}\right]^{2},\qquad B(r)=\left[1+f(r)\right]^{4},
\end{equation}
in which the special case $f=-GM/2r$ represents the Schwarzschild solution,
with a Newtonian limit $\phi=-GM/r$.
In general, if one chooses $f(r)=-\phi(r)/2$, where $\phi(r)$ is any spherical solution of Poisson equation, it
gives rise, in the limit $c\rightarrow\infty$, to a Newtonian potential $\phi$.
This fact sketches a simple procedure to construct general relativistic extensions of previously known Newtonian
solutions, as shown by several authors
 \cite{1964ApJ...140.1512B,vogtLetelier2005MNRAS,doi:10.1111/j.1365-2966.2009.15967.x,doi:10.1111/j.1365-2966.2010.16863.x,Nguyen}.
Here we first focus on the general relativistic extension of the Hernquist potential,
one of the models obtained in \cite{Nguyen}. Then we choose
$f$ as
\begin{equation}\label{fHernquist}
    f(r)=-\frac{\phi(r)}{2}, \qquad \phi(r) =- \frac{\phi_o}{1+(r/a)},
\end{equation}
where $\phi_o$ and $a$ are positive parameters representing the maximum value of $|\phi|$ (at the center of the spherical
configuration) and a scaling radius, respectively.
Note that this metric describes an asymptotically flat space-time with a
Ricci scalar given by
$$
	R=\frac{4\phi_o a (r+a)^2[a(\phi_o - 1)-r]}{r\left[r+a\left(1-\frac{\phi_o}{2}\right)\right]\left[r+a\left(1+\frac{\phi_o}{2}\right)\right]^5},
$$
from which we note  that there are two singularities,
\begin{equation}\label{singularities}
   \mbox{(i)}\:\:\: r=0,\qquad \mbox{(ii)}\:\:\: r=a\left(\frac{\phi_o}{2}-1\right),
\end{equation}
the second one depending on the free parameters $a$ and $\phi_o$. It is easy to see that, for
$\phi_o \leq 2$, singularity (ii) disappears. Also, it can be shown that, for $0<\phi_o\leq 1$, we have $R<0$
 at any radius (see Fig. \ref{fig:Ricci}). In the particular case $\phi_o=1$, we find $R=-4 a (r+a)^2 (r+a/2)^{-1}(r+3a/2)^{-5}$
which means that both singularities, (i) and (ii), disappear. For all other cases, $\phi_o\neq1$, we find always a singularity at origin, $r=0$.

Energy conditions help us to state the range of values for $\phi_o$ leading to physically realizable configurations. In order to
use such conditions, we need the explicit form of the stress-energy tensor, which can be determined via Einstein field equations.
We find that the non vanishing components of the stress-energy tensor can be written in terms of $f$:
\begin{eqnarray}
&&T^{tt}=\frac{4f^3}{\pi G{\phi_o}^2 a r (1+f)^3(1-f)^2},\label{Ttt-f}\\
&&\nonumber\\
&&T^{rr}=\frac{2f^4}{\pi G {\phi_o}^2 a r(1+f)^9(1-f)},\label{Trr-f}\\
&&\nonumber\\
&&T^{\theta\theta}=T^{\psi\psi}\sin\theta=\frac{f^4}{\pi G {\phi_o}^2 a r^3(1+f)^9(1-f)}.\label{Ttheta-f}
\end{eqnarray}
So, it is easy to state that weak energy condition, $-{T^t}_{t}\geq0$, is satisfied if $\phi_o\geq0$.
Strong energy condition, $T=-{T^t}_{t}+{T^r}_{r}+{T^\theta}_{\theta}+{T^\psi}_{\psi}\geq0$, leads to
$$
\frac{4f^3}{(1+f)^5(1-f)}\geq0,
$$
which requires that $0\leq\phi_o\leq 2$. Dominant energy condition, given by
$$
\left|\frac{{T^r}_{r}}{{T^t}_{t}}\right|\leq1,\qquad \left|\frac{{T^\theta}_{\theta}}{{T^t}_{t}}\right|\leq1,\qquad
\left|\frac{{T^\psi}_{\psi}}{{T^t}_{t}}\right|\leq1,
$$
is satisfied if $\phi_o< 4/3$. In summary, we have to choose the parameter $\phi_o$ so that
\begin{equation}\label{energy-conditions}
    0\leq \phi_o < 4/3,
\end{equation}
in order to fulfill weak, dominant and strong energy conditions. This means that physically realizable configurations
described by  (\ref{isotropicMetric2})-(\ref{fHernquist}) have only one singularity, at the center $r=0$.
We shall see, in Sec. \ref{sec:DF-hernquist}, by analyzing the behavior of the corresponding distribution function,   that
we have to choose $\phi_o\leq 2/3$ in order to obtain a DF well defined for $r>0$.

In section \ref{sec:DF-hipervirial} we show that the same procedure can be performed to obtain a general-relativistic extension
of the hypervirial potentials, as proven by Nguyen and Ligman in 2013 \cite{Nguyen}.

\section{Self gravitation equations for static isotropic distributions of matter}\label{sec:selfgraveqs}

In this section we show a detailed derivation of relations which help us to obtain the DF
describing the configuration associated with the metric of (\ref{isotropicMetric}), (\ref{isotropicMetric2}) and (\ref{fHernquist}).
At first, we shall deal with functions $A(r)$ and $B(r)$ representing asymptotically flat space-times, in general, and then we
consider the particular case in which such functions are given by (\ref{isotropicMetric2}) and (\ref{fHernquist}).

The relation between the stress-energy tensor, $T^{\mu\nu}$, and the
the DF, $\DF(x^\mu,\mom^\nu)$ (here $\mom^\mu=\mathrm{d}x^\mu/\mathrm{d}\tau$ is
the $4$-momentum vector and $\tau$ is the proper time), associated with a self-gravitating system, is given by
\begin{equation}{\label{enmomten}}
		T^{\mu\nu}=\int \mom^\mu \mom^\nu\DF \sqrt{-g}\mathrm{d}^4 \mom
	\end{equation}
	where $g=det(g_{\mu\nu})$ and we choose $\mom^t>0$. The phase-space domain associated with a particle
of rest mass $m$ is determined by the shell condition,
	\begin{equation}{\label{shell}}
		g_{\mu\nu} \mom^\mu \mom^\nu =-m^2,
	\end{equation}
from which we can express $\mom^t$ as a function of the remaining phase-space coordinates: $\mom^t=\mom^t(\mom^i,x^\mu)$.
Additionally, neglecting the effect of gravitational encounters in the system, we demand that $\DF$ must satisfy the collisionless Boltzmann equation
\cite{Kremer},
	\begin{equation}{\label{colbolt}}
	\mom^\mu\dfrac{\partial\DF}{\partial x^\mu}-\Gamma^\lambda_{\mu\nu} \mom^\mu \mom^\nu \dfrac{\partial\DF}{\partial \mom^\lambda}=0.
	\end{equation}
Such DF, through relation (\ref{enmomten}) and the Einstein field equations,
$R_{\mu\nu}-g_{\mu\nu}R/2=-8\pi G T_{\mu\nu}$, determines the space-time geometry by the set of relations
\begin{equation}{\label{selfGravityGen}}
R g^{\mu\nu}-2R^{\mu\nu}=16\pi G\int \mom^\mu \mom^\nu\DF \sqrt{-g}\mathrm{d}^4 \mom,
\end{equation}
which we denote here as the \emph{self-gravitation equations}, in the sense that they define, in a self-consistent fashion
(obeying simultaneously Einstein's equations and collisionless Boltzmann equation, or, equivalently, the Einstein-Vlasov system), the evolution of the
system.

Relation (\ref{colbolt})  is equivalent to demand that $\mathrm{d}\DF/\mathrm{d}\tau=0$ \cite{Straumann}, i.e. $\DF$ can be regarded
as an integral of motion.
If the system is endowed by spherical symmetry (or cylindrical or any other) the Jeans theorems guarantee that
 $\DF$ can be expressed as a function the other integrals, which, for the spherical case, are the general relativistic extensions
 of energy $E$ and angular momentum $\mathbf{L}$. In this paper we are focusing on this case.

Motion of free falling test particles in the static isotropic space-time described by (\ref{isotropicMetric})
have one constant of motion, the rest mass $m$, and
 three integrals of motion.
The first of them, an energy-like integral of motion, is the  $t$-component of the
covariant 4-momentum vector, $\mom_t$. The second one is the
azimuthal angular momentum like integral, $\mom_\psi$, and the third one is the general relativistic version of the
total angular momentum, $\sqrt{\mom_\theta ^2+{\mom_\psi ^2}/{\sin^2\theta}}$.
For the sake of simplicity, we adopt the notation
\begin{equation}\label{notationIntegrals}
    \mom_{t}=-E, \qquad \mom_{\psi}=L_z, \qquad \mom_\theta ^2+\frac{\mom_\psi ^2}{\sin^2\theta}=L^2,
\end{equation}
and equations of motion for a
free falling test particle can be cast as
\begin{subequations}\label{motion}
\begin{eqnarray}
m\frac{dt}{d\tau}=\mom^{t}=\frac{E}{A(r)},\qquad\qquad\qquad\qquad\qquad\qquad\:\:\:\:\:\\
 \nonumber\\
 m\frac{d\psi}{d\tau}=\mom^{\psi}=\frac{L_z}{r^{2} B(r)\sin^{2}\theta}, \qquad\qquad\qquad\qquad\:\:\:\:\:\\
  \nonumber\\
m\frac{d\theta}{d\tau}=\mom^\theta=\pm \frac{1}{r^{2} B(r)}\sqrt{L^2-\frac{L_z^2}{\sin^2\theta}},\qquad\qquad\:\:\:\:\:\\
 \nonumber\\
m\frac{dr}{d\tau}=\mom^r=\pm
\sqrt{\frac{E^2}{A(r) B(r)}-\frac{L^{2}}{r^{2}B^{2}(r)}-\frac{m^{2}}{B(r)}},\:\:\:\:
\end{eqnarray}
\end{subequations}
remembering that  phase space coordinates are constrained by the shell condition.
Thus, equations (\ref{shell}) and (\ref{motion}) will be the base for constructing the distribution function.

\subsection{The self-gravitation equations}

Since $g_{\mu\nu}$ does not depend on 4-momentum, equation (\ref{enmomten}) can be written as
	$$
		T^{\mu\nu}=\sqrt{-g} \int \mom^\mu \mom^\nu  \DF\mathrm{d}^4 \mom,
	$$
where the integral is defined in all the phase space domain where $\DF>0$. Since we are dealing with a DF that is function
of the integrals of motion, $E$, $L_z$, $L$ and $m$ (which, through the shell condition (\ref{shell}), can be interpreted as
 an integral of motion), it is convenient to make a transformation from
coordinates $(\mom^{t},\mom^{r},\mom^{\theta},\mom^{\psi})$
to coordinates $(m,E,L_z,L)$. At this point we must be careful
with the transformations of $\mom^{r}$ and $\mom^{\theta}$ since,
according to (\ref{motion}-c) and (\ref{motion}-d), they have two forms, one for each choosing of sign.
Thus, we write
\begin{equation}\label{Prmasmenos}
    \mom^{r}_{+}=\sqrt{\frac{L_m ^2 (r)-L^2}{r^{2}B^{2}(r)}}, \qquad \mom^{r}_{-}=-\mom^{r}_{+},
\end{equation}
where
\begin{equation}\label{Lm}
    L_m (r)=\sqrt{B(r)r^{2}\left(\frac{E^{2}}{A^{2}(r)}-m^{2}\right)},
\end{equation}
and
\begin{equation}\label{Ptetamasmenos}
\mom^\theta_{+}=\frac{1}{r^{2} B(r)}\sqrt{L^2-\frac{L_z^2}{\sin^2\theta}},\qquad
\mom^{\theta}_{-}=-\mom^{\theta}_{+}.
\end{equation}
Therefore we have to write
	\begin{multline*}
	T^{\mu\nu}=\sqrt{-g} \left[\int \mom^\mu \mom^\nu \DF\mathrm{d} \mom^{t}\mathrm{d} \mom^{r}_{+}\mathrm{d} \mom^{\theta}_{+}\mathrm{d}
 \mom^{\psi}\right.\\+ \int \mom^\mu \mom^\nu  \DF\mathrm{d} \mom^{t}\mathrm{d} \mom^{r}_{-}\mathrm{d} \mom^{\theta}_{+}\mathrm{d}
  \mom^{\psi}\\+ \int \mom^\mu \mom^\nu  \DF\mathrm{d} \mom^{t}\mathrm{d} \mom^{r}_{+}\mathrm{d} \mom^{\theta}_{-}\mathrm{d}
   \mom^{\psi}\\+ \left.\int \mom^\mu \mom^\nu \DF\mathrm{d} \mom^{t}\mathrm{d} \mom^{r}_{-}\mathrm{d} \mom^{\theta}_{-}\mathrm{d}
   \mom^{\psi}\right].
	\end{multline*}
In particular, the expression for components $T^{rr}$ and $T^{\theta\theta}$ requires a replacement of $\mom^{r}$ and $\mom^{\theta}$ by
$\mom^{r}_{+}$, $\mom^{r}_{-}$, $\mom^{\theta}_{+}$ and$/$or $\mom^{\theta}_{-}$, according to the variables of integration. For example,
in the above expression, the term involving $\mathrm{d} \mom^{r}_{+}\mathrm{d} \mom^{\theta}_{-}$ requires that we set
$\mom^{r}\rightarrow\mom^{r}_{+}$, when calculating
$T^{rr}$, and it will require $\mom^{\theta}\rightarrow\mom^{\theta}_{-}$, when computing $T^{\theta\theta}$.
Note that in all cases, the Jacobian of the transformation is
$$
	 \left|\dfrac{\partial(\mom^t,\mom^r,\mom^\theta,\mom^\psi)}{\partial(m,E,L,L_z)}\right|=
\frac{mL}{\mom^{r}_{+}\mom^{\theta}_{+}AB^{4}r^{6}\sin^{2}\theta},
	$$
and the domain of integration is given by the relations
	\begin{equation}{\label{limits}}
	\begin{cases}
	-L\sin\theta\leq L_z\leq L\sin\theta\\0\leq L\leq L_m,\\
	m\sqrt{A}\leq E\leq m, \\
	0\leq m\leq\infty.
	\end{cases}
	\end{equation}
The bounds for $E$ arise from the shell condition and from the escape energy, which
can be elucidated from relation (\ref{motion}-d). At $r\rightarrow \infty$ we have $A=B=1$,
since we are assuming that (\ref{isotropicMetric}) represents an asymptotically flat metric,
and we have
	$$
	|\mom^r|=\sqrt{{E^2-m^2}},\qquad r\rightarrow \infty
	$$
Then the escape energy, at $r\rightarrow \infty$, is $E=m$ (remember that
we chose energy to be positive), corresponding to the value $|\mom^r|=0$.	
Thus, we can state that particles with energy larger than $m$ can not belong to the configuration.

It can be shown, from (\ref{isotropicMetric}), that components of the stress-energy tensor that could be
non-vanishing are
$T^{tt}$, $T^{rr}$, $T^{\theta\theta}$ and $T^{\psi\psi}$, whereas the other components vanish in any case (i.e. for
 an arbitrary DF). This fact can be checked directly from (\ref{enmomten}), except for the case of $T^{t\psi}$, which
does not vanish trivially. However,
since the stress-energy
tensor is a function only of radius $r$, it is required that
the DF has the form
$$
\DF(m,E,L,L_z)=\DF(m,E,L),
$$
leading to $T^{t\psi}=0$
and simplified expressions for the non-vanishing components:
$$
T^{tt}=\frac{4\pi }{r^{2}A^{5/2}B^{3/2}} \int\limits_{0}^{\infty}\int\limits_{m\sqrt{A}}^{m}\int\limits^{L_m}_{0}
\frac{E^2m L\DF }{\mom^{r}_+}\mathrm{d} L\mathrm{d} E\mathrm{d}m,
$$
$$
T^{rr}=\frac{4\pi }{r^{2}A^{1/2}B^{3/2}}
\int\limits_{0}^{\infty}\int\limits_{m\sqrt{A}}^{m}\int\limits^{L_m}_{0}
{{\mom^r_{+}}mL\DF }\mathrm{d} L\mathrm{d} E\mathrm{d}m,
$$
$$
T^{\theta\theta}=\frac{2\pi }{r^{6}A^{1/2}B^{7/2}}\int\limits_{0}^{\infty}\int\limits_{m\sqrt{A}}^{m}\int\limits^{L_m}_{0}
\frac{mL^3\DF }{\mom^{r}_+}\mathrm{d} L\mathrm{d} E\mathrm{d}m,
$$
and $T^{\psi\psi}=T^{\theta\theta}/\sin^2\theta$.
In many applications it is common to assume that
the mass for every constituent of the system is the same. This lead us to
replace $\DF(m,E,L)$ by $\DF(E,L)$, which now satisfies the following simplified form:
\begin{equation}\label{tt-2}
T^{tt}=\frac{4\pi m}{r^{2}A^{5/2}B^{3/2}}\int\limits_{m\sqrt{A}}^{m}\int\limits^{L_m}_{0}
\frac{E^2 L\DF(E,L) }{\mom^{r}_+}\mathrm{d} L\mathrm{d} E,\:
\end{equation}
\begin{equation}\label{tr-2}
T^{rr}=\frac{4\pi m}{r^{2}A^{1/2}B^{3/2}}
\int\limits_{m\sqrt{A}}^{m}\int\limits^{L_m}_{0}
{{\mom^r_{+}}L\DF(E,L) }\mathrm{d} L\mathrm{d} E,
\end{equation}
\begin{equation}\label{tte-2}
T^{\theta\theta}=\frac{2\pi m}{r^{6}A^{1/2}B^{7/2}}\int\limits_{m\sqrt{A}}^{m}\int\limits^{L_m}_{0}
\frac{L^3\DF(E,L) }{\mom^{r}_+}\mathrm{d} L\mathrm{d} E.\:\:
\end{equation}
The above relations, remembering that $T_{\mu\nu}=[g_{\mu\nu}(R/2)-R_{\mu\nu}]/(8\pi G )$,
can be regarded as the self-gravitation equations in the case of a general static isotropic metric.
Then, by defining the functions $A$ and $B$ in eq. (\ref{isotropicMetric}), in principle, we can determine $\DF(E,L)$
through equations (\ref{tt-2}), (\ref{tr-2}) and (\ref{tte-2}). A similar expression is shown in \cite{Fackerell_1968} for a metric in the standard form.

\subsection{Models with $P_{\theta}=kP_{r}$}\label{sec:anisotropic}

In this section we assume that the configuration can be regarded as a fluid with a dynamics described in terms of
the energy density $\rho$, the radial pressure $P_{r}$ and the tangential pressure $P_{\theta}$ (or $P_{\varphi}$). In this context it is
useful to distinguish between isotropic ($P_{r}=P_{\theta}$) and anisotropic systems ($P_{r}\neq P_{\theta}$), by introducing the
anisotropy parameter
\begin{equation}\label{param-anisotrop}
\beta=1-\frac{P_{\theta}}{P_{r}}.
\end{equation}
Thus, isotropic fluids are represented by $\beta=0$ and anisotropic fluids are characterized by a function $\beta(r)$ which, in general, does not vanish.
Here we focus in the case in which the anisotropy parameter is a real constant, $\beta=1-k$, i.e. fluids such that $P_{\theta}=kP_{r}$. We
will show that this particular class of systems with constant anisotropy are characterized by a distribution function of the form
$\DF=\xi(E)L^{2\left(k-1\right)}$.

At first, remember that $\rho$, $P_{r}$ and $P_{\theta}$ are related with
the stress-energy tensor by the relations
$$
\rho=-{T^{t}}_{t}, \qquad P_{r}={T^{r}}_{r}, \qquad P_{\theta}={T^{\theta}}_{\theta}={T^{\varphi}}_{\varphi},
$$
which, by using (\ref{tt-2}), (\ref{tr-2}) and (\ref{tte-2}), can be written as 	
	\begin{equation}\label{rho1}
	\rho=\frac{4\pi m}{r^{2}(BA)^{\frac{3}{2}}} \int\limits_{m\sqrt{A}}^{m}\int\limits^{L_m}_{0}
	\frac{E^2 L\DF (E,L)}{\mom^{r}_+}\mathrm{d} L\mathrm{d} E,
	\end{equation}
	\begin{equation}\label{Pr1}
	P_{r}=\frac{4\pi m}{r^{2}\sqrt{BA}}
	\int\limits_{m\sqrt{A}}^{m}\int\limits^{L_m}_{0}
	{{\mom^r_{+}}L\DF (E,L)}\mathrm{d} L\mathrm{d} E,
	\end{equation}
	\begin{equation}\label{Ptheta1}
	P_{\theta}=\frac{2\pi m }{r^{4}B^{\frac{5}{2}}\sqrt{A}}   \int\limits_{m\sqrt{A}}^{m}\int\limits^{L_m}_{0}
	\frac{L^3\DF (E,L)}{\mom^{r}_+}\mathrm{d} L\mathrm{d} E.
	\end{equation}
Note that, by choosing $\DF(E,L)=\xi(E)L^{2\left(k-1\right)}$ (with $k$ a constant) in the above equations we can write
$P_{\theta}=kP_{r}$. Also we can prove that by setting $P_{\theta}=kP_{r}$, then the DF, necessarily, must have the form
$\xi(E)L^{2\left(k-1\right)}$.

Let us write the statement $P_{\theta}=kP_{r}$ by using (\ref{Pr1})-(\ref{Ptheta1}) and taking into account, for now,
only the integral with respect to $L$ :
 $$		
		\int^{L_m}_{0}
		\frac{L^3\DF }{\sqrt{L_m ^2-L^2}}\mathrm{d} L=
		2k \int^{L_m}_{0}
		L\DF \sqrt{L_m ^2-L^2}\,\mathrm{d} L.
$$
Now, we can		integrate by parts the right hand side of the above expression,
		\begin{multline*}
		2\int\limits^{L_m}_{0}
		L\DF \sqrt{L_m ^2-L^2}\,\mathrm{d} L=\int\limits^{L_m}_{0}
		\frac{L^3\DF}{\sqrt{ L_m ^2-L^2}}\mathrm{d} L-\;\;\;\;\;\;\\\;\;\;\;\;\;\int\limits^{L_m}_{0}
		L^2\sqrt{L_m ^2-L^2}\frac{\partial\DF}{\partial L}\mathrm{d} L-L_{m}\lim\limits_{L\rightarrow 0}\left(L^2\DF\right).
		\end{multline*}
It can be shown that $\lim_{L\rightarrow 0}(L^2\DF)=0$,
for any $\DF(E,L)$ satisfying (\ref{tt-2}), (\ref{tr-2}) and (\ref{tte-2}) (see appendix \ref{appendix2} for a detailed proof).
Then, we can write
		 $$
		 \int\limits^{L_m}_{0}L \sqrt{L_m ^2-L^2}\left[
		 2\left(k-1\right)\DF-L\frac{\partial\DF}{\partial L}\right]\mathrm{d} L=0,
		 $$
		which has to be satisfied for every $L_{m}$ (or for every $E$). Therefore,
		$$
		2\left(k-1\right)\DF-L\frac{\partial\DF}{\partial L}=0\Rightarrow\DF=\xi(E)L^{2\left(k-1\right)}.
		$$

Finally, we can state the following theorem:

\begin{theorem}\label{Teorema1}
	Let $k$ be a constant and $\DF$ a distribution function that satisfies the self-gravitation equations for static spherically symmetric
configurations. Then $P_{\theta}=k P_{r}$ if and only if $\DF(E,L)=\xi(E)L^{2\left(k-1\right)}$.
\end{theorem}

Thus, models with constant anisotropy $\beta$ are characterized by a distribution function proportional to $\xi(E)L^{-2\beta}$.
In the next sections we show that the Hernquist model, as well as the so-called hipervirial models, belong to this class of systems. 	
		
\section{Distribution function for general-relativistic Hernquist model}\label{sec:DF-hernquist}

Here we show how to derive a relativistic DF for a relativistic Hernquist model, given by (\ref{isotropicMetric2})-(\ref{fHernquist})
by using  the self-gravitation equations (\ref{tt-2})-(\ref{tte-2}). Since the factor $\sqrt{A}$
appears repeatedly in eqs. (\ref{tt-2})-(\ref{tte-2}),
it is important to note that (\ref{isotropicMetric2})-(\ref{fHernquist}) imply
$$
f(r)=
\begin{cases}
		\frac{1-\sqrt{A(r)}}{1+\sqrt{A(r)}}, \qquad r>a(\frac{\phi_o}{2}-1)\\\\
		\frac{1+\sqrt{A(r)}}{1-\sqrt{A(r)}}, \qquad 0<r\leq a(\frac{\phi_o}{2}-1)
\end{cases}
$$
Since energy conditions require that $0\leq \phi_o < 4/3$ (remember relation (\ref{energy-conditions})) we find that
$a[(\phi_o/2)-1]<-a/3$, which imply  two facts: (i) there are not values for $r$ satisfying $ 0<r\leq a[(\phi_o/2)-1]$; (ii)
all the (positive) values for $r$ satisfy $r>a[(\phi_o/2)-1]$. Therefore, the only option for $f$, consistent with all the
energy conditions, is
\begin{equation}\label{fonly}
    f(r)=\frac{1-\sqrt{A(r)}}{1+\sqrt{A(r)}}, \qquad r>0.
\end{equation}
This means that relations (\ref{Ttt-f})-(\ref{Ttheta-f}), by introducing (\ref{fonly}),  can now be rewritten as
\begin{eqnarray}
&&T^{tt}=\frac{(1-\sqrt{A})^{3}(1+\sqrt{A})^2}{2^3\pi G{\phi_o}^2 a r A},\label{Ttt-gtt}\\
&&\nonumber\\
&&T^{\theta\theta}=\frac{T^{rr}}{2r^2}=\frac{(1-\sqrt{A})^4(1+\sqrt{A})^6}{2^{10}\pi G r^3{\phi_o}^2 a\sqrt{A}},\label{Trr-gtt}
\end{eqnarray}
This form is particularly useful when compared with the corresponding equations obtained from (\ref{tt-2}), (\ref{tr-2}) and (\ref{tte-2}).
Indeed  we find that $P_{\theta}=P_{r}/2$.  By using the result of Proposition \ref{Teorema1}, this fact implies that
$$
\DF(E,L)=\xi(E)L^{-1},
$$
where $\xi(E)$ is a function to be found by comparing  the right hand side of eqs. (\ref{Ttt-gtt}), (\ref{Trr-gtt})
with the right hand side of (\ref{tt-2})-(\ref{tte-2}).
After some calculations we obtain two relations for $\xi$:
		\begin{equation}{\label{int1}}
		\int\limits_{m\sqrt{A}}^{m}E^2\xi(E)\mathrm{d} E=\frac{A^{3/2}(1-\sqrt{A})^3}{2^2 \pi^3 m G {\phi_o}^2 a },
		\end{equation}
		\begin{equation}{\label{int2}}
		\int\limits_{m\sqrt{A}}^{m}\xi(E)\left[E^2-m^2 A\right]\mathrm{d} E=\frac{A(1-\sqrt{A})^4}{2^{3}\pi^3 m G{\phi_o}^2a} .
		\end{equation}
From (\ref{int1}) we find
$$
\xi(E)=\frac{3}{4 m^4\pi^3G{\phi_o}^2a}\left(\frac{2E}{m}-1\right)\left(\frac{E}{m}-1\right)^2,
$$
which is consistent with relation (\ref{int2}).

\begin{figure}[ht]
 \includegraphics[scale=0.6]{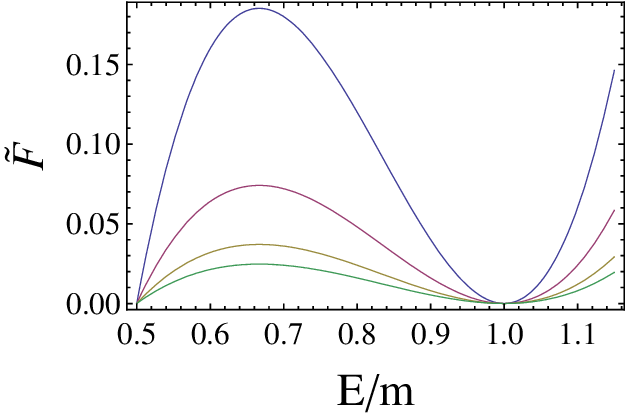}
  \caption{Dimensionless DF, $\tilde{F}=\xi_{o}^{-1} \DF$, for the general relativistic extension of Hernquist potential
as a function of $E/m$, for different values of $L/m$: 0.2 (blue), 0.5 (violet), 1 (yellow), 1.5 (green).
  }
  \label{fig:DF-Hernquist}
\end{figure}

For the sake of simplicity, we introduce the dimensionless energy $\varepsilon$ and the dimensionless angular momentum
$l$, as
\begin{equation}\label{ELdimensionless}
\varepsilon\equiv E/m,\qquad l\equiv L/m,
\end{equation}
and thus we can write the explicit analytic form of the
DF corresponding to the general relativistic extension of Hernquist model, as a function of $\varepsilon$ and $l$:
		\begin{equation}{\label{hdf}}
		\DF(\varepsilon,l)=\xi_{o}l^{-1}\left(2\varepsilon-1\right)\left(\varepsilon-1\right)^2,
		\end{equation}
with
		\begin{equation}{\label{hdfo}}
		\xi_{o}=3\left(4 m^5\pi^3G{\phi_o}^2a\right)^{-1}.
		\end{equation}
Note that such DF is negative for $E<m/2$, so, in principle, we would have to restrict its domain
to values of energy larger than $m/2$. In the next section we show that a natural way to do this is by constraining
the values of the free parameter $\phi_o$. In figure \ref{fig:DF-Hernquist} we plot the behavior of the DF given by (\ref{hdf}),
once $\phi_o$ has been chosen adequately.

  \subsection*{Constraining the values for  $\phi_o$ }\label{sec:phio}

Self-gravitation equations (\ref{tt-2})-(\ref{tte-2})
impose some restrictions to the stress-energy tensor (not necessarily equivalent to energy conditions), when one
demands that $\DF\geq 0$.
They can be summarized as,
\begin{subequations}{\label{restr}}
	\begin{align}
	T_{\mu\nu} &\geq 0,\\
	\nonumber\\
	T&\leq 0,
	\end{align}
\end{subequations}
Indeed these  restrictions are stronger  than the weak, null, dominant and strong energy conditions.
When they are applied to the stress-energy tensor given by
(\ref{Ttt-f})-(\ref{Ttheta-f}),
we find the following inequality
\begin{equation}{\label{frest}}
0\leq f\leq 1/2,
\end{equation}
which in terms of the radial coordinate $r$ is equivalent to state that
$$
r\geq a\left(\phi_o-1\right).
$$
This means that a real, positive DF, determining the stress-energy tensor could
be well defined	only for $r\geq a\left(\phi_o-1\right)$. So, the maximum value of $\phi_o$	that
permits a DF well defined at the entire configuration space, $r\geq0$, is $\phi_o=1$.

The  bounding value for $\phi_o$ can be diminished by taking into account that
the DF of eq. (\ref{hdf}) is negative for
$E<m/2$ and remembering that the minimum value for
a particle's energy is $E_{\text{min}}=m\sqrt{A}$. Therefore, situations where
$\sqrt{A}<1/2$, which in this case equals to state that $r<a(3\phi_{0}/2 -1)$, are
not described for a positive DF given by (\ref{hdf}). Such a DF only could describe situations
where

		$$
		r\geq a\left(\frac{3\phi_{0}}{2}-1\right),
		$$
which means that,  $\phi_o= 2/3$ is now the maximum value for $\phi_o$ such that $\DF$ is positive and well defined for $r\geq0$.
By choosing this bound for $\phi_o$ we guarantee that $E\geq m/2$ for all situations. Thus, finally we can state that the
set of values for the free parameter $\phi_o$ are given by
\begin{equation}\label{phi2-3}
 \boxed{\:\:   0\leq\phi_o\leq 2/3 \:\:}\:\:,
\end{equation}
in order to obtain a self-consistent relativistic Hernquist model, charaterized by a DF well defined at the entire configuration space.

\section{Distribution Functions for a general-relativistic version of the Hypervirial family}\label{sec:DF-hipervirial}

The formalism used in the preceding sections
can also be applied in the case of the hypervirial family,
to which  Hernquist model belongs. In Newtonian gravity, the hipervirial potentials are given by
\begin{equation}\label{phi-hipervirial}
    \phi_{n}(r)=-\frac{\phi_{no}}{\left[1+\left(r/a\right)^n\right]^{\frac{1}{n}}},
\end{equation}
where $n$ is a positive integer and $\phi_{no}$, $a$  positive real constants. Each member
is characterized by a DF proportional to $E^{(3n+1)/2}L^{n-2}$.

As in the case of Hernquist model (the particular case $n=1$ of
(\ref{phi-hipervirial})),
a physically reasonable relativistic extension, introduced previously in \cite{Nguyen}, is performed by defining
$f=-\phi_{n}/2$ in relation (\ref{isotropicMetric2}), leading to a stress-energy tensor of the form
\begin{eqnarray}
    T^{rr}&=&\frac{2^{2n-1}f^{2n+2}}{\pi a^{n} {\phi_{no}}^{2 n}G r^{2-n} (1-f)(1+f)^9}=\frac{2r^2}{n}T^{\theta\theta}\nonumber\\
    &&\nonumber\\
    &=&\frac{f(1-f)}{(n+1)(1+f)^6}T^{tt},\label{T-hipervirial1}
\end{eqnarray}
and $T^{\mu\nu}=0$ for $\mu\neq\nu$.
From \eqref{T-hipervirial1} is easy to see that $P_{\theta}=(n/2)P_{r}$, which, by using Proposition \ref{Teorema1}, implies that the
corresponding DF can be written as
$$
\DF=\xi(E)L^{n-2}, \qquad n=1,2,...
$$
By introducing the above expression  into (\ref{tt-2})-(\ref{tte-2}) we obtain
$$
\int\limits_{m\sqrt{A}}^{m}{\xi^{*}(E)E^2}\left(\frac{E^2}{A}-m^2\right)^{\frac{n-1}{2}}
\mathrm{d}E=2A^{3/2}\left(1-\sqrt{A}\right)^{2n+1},
$$

$$
\int\limits_{m\sqrt{A}}^{m}\xi^{*}(E)\left(\frac{E^2}{A}-m^2\right)^{\frac{n+1}{2}} \mathrm{d}E=\left(1-\sqrt{A}\right)^{2n+2},
$$
where
$$
\xi(E)=\xi^{*}(E)\frac{\left(n+1\right) \Gamma \left(\frac{n+1}{2}\right)}{2^{4}\pi^{\frac{5}{2}} a^{n} {\phi_{no}}^{2 n} G \Gamma \left(\frac{n}{2}\right) m}.
$$
These two relations are essentially the same: the first one can be obtained by taking the derivative of the second one with respect to
$\sqrt{A}$.
So, in this case, we can choose the second one relation (the simpler one)
as the integral equation to be solved, in order to find an explicit expression for function $\xi$. Here, for simplicity, we define
$\sqrt{A}=x$, which leads to

\begin{equation}\label{integral-eq-hiperv}
    \int\limits_{m x}^{m}\xi^{*}(E)\left[{{\left(\frac{E}{m}\right)}^2}-{x}^2\right]^{\frac{n+1}{2}}
 \mathrm{d}E=\left(\frac{x}{m}\right)^{n+1}\left(1-x\right)^{2n+2}.
\end{equation}
In order to solve the above relation it is convenient to consider, separately, two cases: (i) $n=1,3,5,...$
and (ii) $n=0,2,4,...$. Each of these options will lead to two kinds of distribution
functions.

\begin{enumerate}
\item[(i)]
By choosing $n=2p+1$, for $p=0,1,2,\ldots$, in eq. (\ref{integral-eq-hiperv}), we find that
$$
\xi_{2p+1}(E)=
\sum\limits_{k=1}^{4p+4}a_{2p+1}\left(\frac{E}{m}\right)^{k-1},
$$
where the $a_{2p+1}$ are constants that will be specified
later (see eq. (\ref{ank})). Note that the DF corresponding to the relativistic extension of Hernquis model is obtained for $p=0$ (or $n=1$).
The next case, $p=1$ (or $n=3$) is described by a function
$$
\xi_3(E)\propto\left(1-\frac{E}{m}\right)^5 \left[40 \left(\frac{E}{m}\right)^2-31 \frac{E}{m}+5\right],
$$
which must be restricted to a domain given (approximately) by
$0\leq E/m\leq 0.2289$ and $0.5461\leq E/m\leq 1$, in order to have a positive DF. For the other cases,
$p=2,3,..$ the function $\xi$ also can be written in the form $\xi_{2p+1}\propto (1-\varepsilon)^{3p+2}g(\varepsilon)$, where
$g$ is a polynomial of degree $p+1$ in $\varepsilon$.

\item[(ii)] The case in which $n$ is even, i.e. $n=2p$ for $p=0,1,2,\ldots$ in equation (\ref{integral-eq-hiperv}), demands a little more attention.
By computing the derivative in $x$ of (\ref{integral-eq-hiperv}), $p+1$ times,  we have
$$
\int\limits_{mx}^{m}\frac{\xi(E)\mathrm{d}E}{\sqrt{\left(E/m\right)^2-x^2}}\propto\\
\left(-\frac{1}{x}\frac{\mathrm{d}}{\mathrm{d}x}\right)^{p+1}\left[\frac{x^{2p+1}\left(1-x\right)^{4p+2}}{m^{2p+1}(2p+1)!!}\right].
$$
Note that
the right side has the form of an Abel integral, so the function $\xi$ can be determined explicitly by performing
the Abel transformation. Thus, after some calculations we find
$$
{\xi_{2p}}(E)=\frac{2E}{m}\sum\limits_{k=0}^{4p+2}b_{2pk}
\int\limits_{\frac{E}{m}}^{1}\frac{x^{k-2}}{\sqrt{x^2-\left(\frac{E}{m}\right)^2}}\mathrm{d}x.
$$
where $b_{2pk}$ are constants given by relations (\ref{bnk}).
For example, the case $p=1$ (or $n=2$), for which the $L$-dependence is dropped,
lead us to the DF corresponding to the relativistic extension of the Plummer model:
$$
\begin{array}{cc}
 & \xi_{2}\propto E^{-1}\sqrt{1-\frac{E^2}{m^2}}+\frac{8 }{4\pi} \sqrt{1-\frac{E^2}{m^2}}
\left( \frac{1733 E^3}{m^7}+\frac{1274E}{m^{5}}\right)\\
 &  \\
 & -\frac{15}{4 \pi }\left( \frac{21 E^5}{m^9}+
\frac{140 E^3}{m^7}+ \frac{40 E}{m^5}\right) \ln\left(\frac{m}{E}\sqrt{1-\frac{E^2}{m^2}}+\frac{m}{E}\right)
\end{array}
$$
\end{enumerate}

We can summarize our results through the following relations
\begin{equation}\label{eq:hypervirialodddf}
\DF_{n}^{(\text{odd})}=l^{n-2}\sum\limits_{k=1}^{2n+2}a_{nk}\epsilon^{k-1},\qquad n=1,3,5,...
\end{equation}
where
\begin{eqnarray}
a_{nk}&=&\binom{2n+2}{k}\frac{(-1)^{k+\frac{n-1}{2}} (k+n+1)!! k}{2^4 \pi^3 a^n {\phi_{no}}^{2n}Gm^5\Gamma\left(\frac{n}{2}\right)k!!(n+1)!!}\nonumber\\
&&\nonumber\\
&\times&(n+1)\sqrt{\pi}\Gamma\left(\frac{n+1}{2}\right),\:\: n=1,3,5,..,\label{ank}
\end{eqnarray}
for DFs with odd index, and
\begin{equation}\label{eq:hypervirialevendf}
\DF_{n}^{(\text{even})}=l^{n-2}\epsilon\sum\limits_{k=0}^{2n+2}b_{nk}\int\limits_{\epsilon}^{1}\frac{x^{k-2}\mathrm{d}x}{\sqrt{x^2-\epsilon^2}},
\qquad n=0,2,4,...,
\end{equation}
where
$$
b_{n0}=\frac{(-1)^{1-\frac{n}{2}}(n+1)\Gamma\left(\frac{n+1}{2}\right)}{2^3 \pi^3 a^n {\phi_{no}}^{2n}Gm^5\sqrt{\pi}\Gamma\left(\frac{n}{2}\right)},
\qquad b_{n1}=0,
$$
\begin{eqnarray}
b_{nk}&=&\binom{2n+2}{k}\frac{{(k+n+1)!!}(-1)^{k-\frac{n}{2}}(k-1)}{2^3 \pi^3 a^n {\phi_{no}}^{2n}Gm^5(k-1)!!(n+1)!!}\nonumber\\
&&\nonumber\\
&\times&\frac{(n+1)\Gamma\left(\frac{n+1}{2}\right)}{\sqrt{\pi}\Gamma\left(\frac{n}{2}\right)},\:\:\;k\geq2,\:\: n=0,2,4,...\label{bnk}
\end{eqnarray}

\begin{figure*}[ht]
 \includegraphics[scale=0.5]{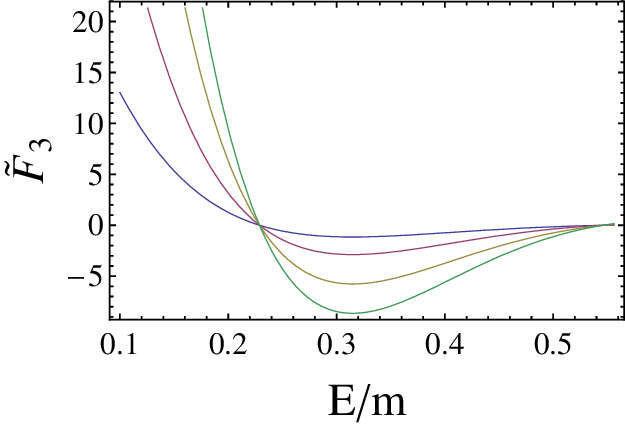}\includegraphics[scale=0.5]{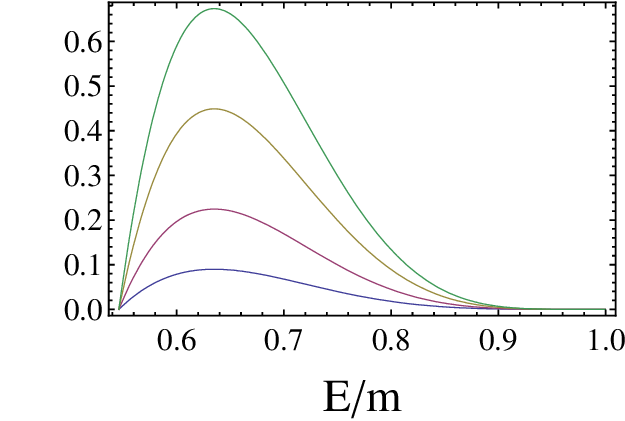}\includegraphics[scale=0.5]{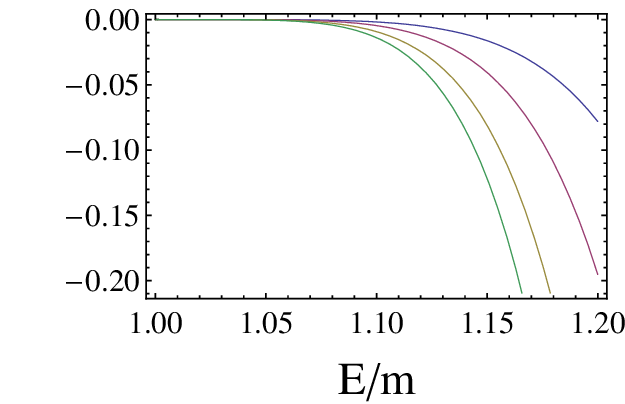}
  \caption{Dimensionless DF corresponding to the hypervirial model $n=3$, for different values of $L/m$: 0.2 (blue), 0.5 (violet), 1 (yellow), 1.5 (green).
  This DF is positive for $0\leq E/m\leq 0.2289$ (left) and for $0.5461\leq E/m\leq 1$ (central panel). Note that probability density
  reaches higher values in the first range. For $E/m>1$, this DF has negative values (right panel).}
  \label{fig:DF-n3}
\end{figure*}

\begin{figure*}[ht]
 \includegraphics[scale=0.5]{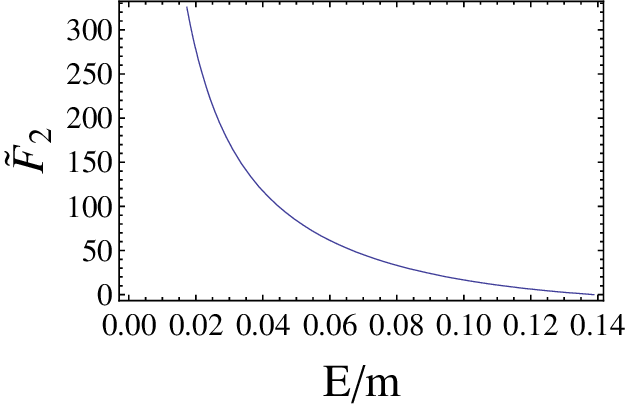}\includegraphics[scale=0.5]{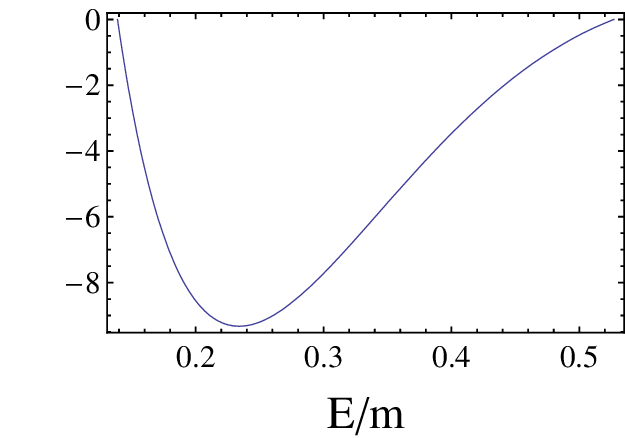}\includegraphics[scale=0.5]{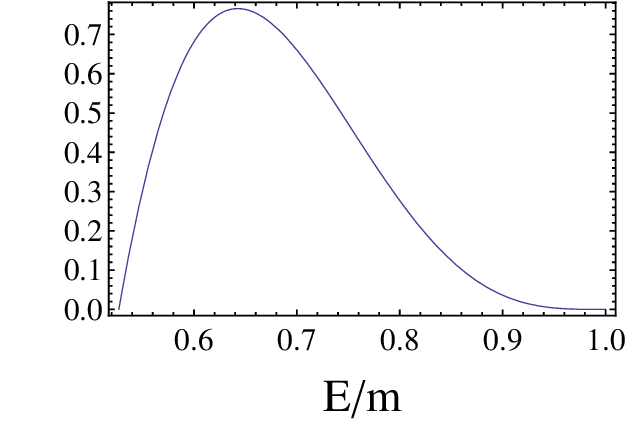}
  \caption{Dimensionless DF corresponding to the hypervirial model $n=2$, which is a relativistic extension of Plummer model. The DF
  is positive for $0\leq E/m \leq 0.1388$ (left panel) and $0.5270\leq E/m < 1$ (right panel) and is negative for $0.1388< E/m < 0.5270$
  (central panel). }
  \label{fig:DF-n3}
\end{figure*}

\begin{figure}[ht]
	\includegraphics[scale=0.6]{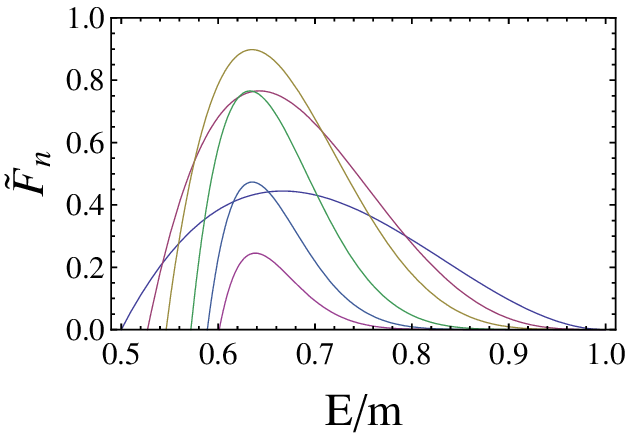}
	\caption{
		Dimensionless DF, $\tilde{F}_{n}= 2^5 m^5\pi^3 G {\phi}^{2n}_{no} a^{n} \DF_{n}$, for the general relativistic extension of the Hypervirial family
		as a function of $E/m$ with $L/m=2$, for different values of $n$: n=1 (red), n=2 (Blue), n=3(Green), n=5 (Orange), n=7 (Purple), n=9 (Brown). }
	\label{fig:HyperviralFamily}
\end{figure}

\begin{table}
	\centering
	\caption{Upper bound value of $\phi_{no}$ for different $n$.} \label{table1}
	\begin{tabular}{| c | c |}
			\hline
		$\quad$ n $\quad$& $\phi_{no}$ \\
		\hline
		$\quad$1$\quad$ & 2/3 \\$\quad$2$\quad$ & 0.619472 \\$\quad$3$\quad$& 0.587143 \\$\quad$5$\quad$& 0.544734 \\$\quad$7$\quad$& 0.517533 \\$\quad$9$\quad$& 0.498276\\
		\hline
	\end{tabular}
\end{table}

As done in Sec. \ref{sec:phio}, we can choose the values of $\phi_{no}$ so that $\DF_n$ be positive everywhere in configuration space, $r>0$.
Figure \ref{fig:HyperviralFamily} suggests that the upper bound for $\phi_{no}$ decreases with $n$, as confirmed by the values of Table \ref{table1}.

\section{Conclusion}

We derived an analytic expression for the DF corresponding to the
general relativistic extension of the Hernquist model presented in \cite{Nguyen}.
In the derivation we considered the self-gravitating equations for asymptotically flat static isotropic space-times, from which
we established that anisotropic models so that $P_{\theta}=k P_{r}$, with $k$ constant, are characterized
 by a DF of the form $\DF=\xi(E)L^{2\left(k-1\right)}$  (proposition \ref{Teorema1}). For the Hernquist case, corresponding to $k=1/2$,
 we find $\DF(E,L)\propto L^{-1}\left(2{E}/{m}-1\right)\left(1-{E}/{m}\right)^2$, from which we established that the upper bound of free parameter
 $\phi_o$ is $2/3$ (lesser than the one obtained in \cite{Nguyen}), in order to have a DF defined at the entire configuration space, $r>0$.

Exploiting our experience with the Hernquist potential we also derived analytic expressions for the DF of the Hypervirial family,
which satisfies $P_{\theta}=(n/2) P_{r}$ for the $n$th member  (Hernquist model is the first member, $n=1$).
Proposition \ref{Teorema1} implies that the DF corresponding to the $n$th member
is of the form $\DF_{n}=\xi_{n}(E)L^{2-n}$, where we have to distinguish between odd and even values of $n$,
in order to encompass in a simple fashion  all cases (eqs. (\ref{eq:hypervirialodddf}) and (\ref{eq:hypervirialevendf})). Thus we
find two subfamilies in the set of hypervirial models, which now can be regarded as a self-consistent family of
models in the context of general relativity.

We note that the free parameter $\phi_{no} $, corresponding to the $n$th member of the hypervirial family,
has an upper bound which diminishes by increasing $n$. Such upper bound, as in the case of Hernquist model, was
chosen in such a way that the DF was positive for $r>0$. However, one could choose different upper bounds for these parameters
when taking into account a reduced configuration space, for example given by $r\geq r_*$, where $r_*$ is a positive constant. This can
be used to model situations composed by two solutions of Einstein equations, one of them defined in $0<r<r_*$ (the solution inside the region
bounded by the shell $r=r_*$)
and the other one, an Hypervirial solution, defined in $r\geq r_*$.
 In such a case, the DF has to be defined by parts and
junction conditions has to be satisfied in the shell $r=r_*$ (see for example \cite{doi:10.1063/1.1621056}).


\appendix

\section{Hernquist potential in Newtonian Gravity}\label{hpng}

	The distribution function (DF) for the Hernquist potential is given by \cite{Nguyen}		
	\begin{equation}\label{DF}			
		\DF(\varepsilon,L)=A{\varepsilon}^{\beta} L ^ {2\alpha}		
	\end{equation}
	where $L$ is the norm of the specific angular momentum and $\varepsilon=\phi_{*} - E$
is the relative energy (in this case we have to set $\phi_{*}=0$).
	This is the same distribution function used by Nguyen et al.
in order to develop a family of potential-density pairs, including the Hernquist model
	as a particular case. The mass density can be found by integrating the distribution function over the velocity space,
	$$
		\rho=\int \DF(\varepsilon,L) {\mathrm{d}}^3\nu,
	$$
	which, by introducing (\ref{DF}) and using  spherical coordinates, leads to
	\begin{equation}
		\rho=\int_0^{2\pi} \int_0^{\pi} \int_0^{\nu_e}
A{\varepsilon}^{\beta} L ^ {2\alpha} \nu^2 \sin\eta{\mathrm{d}}\nu {\mathrm{d}}\eta {\mathrm{d}}\kappa
	\end{equation}
	where $\nu_e = \sqrt{-2\phi}$ is the escape velocity and  $\phi$ is the gravitational potential.
	Since in spherical coordinates we can write
	$L^2=r^2 {\nu}^2 \sin^2\eta$ and $\varepsilon=-E=-\phi-\nu^2 /2$,  we have
	$$
		\rho=2\pi A \int_0^{\pi} \sin^{2 \alpha+1}\eta {\mathrm{d}}\eta \int_0^{\nu_e}\left(-\dfrac{\nu^2}{2}-\phi\right)^{\beta}
		r^{2 \alpha} \nu^{2\alpha +2 } {\mathrm{d}}\nu
	$$
	The first integral above is basically a constant, so by taking $2\pi A \int_0^{\pi} \sin^{2 \alpha+1}\eta {\mathrm{d}}\eta=B$, we have
	\begin{equation}
		\rho=B  r^{2 \alpha} \int_0^{\nu_e}\left(-\dfrac{\nu^2}{2}-\phi\right)^{\beta} \nu^{2\alpha +2 } {\mathrm{d}}\nu
	\end{equation}
	Now, in order to compute the second integral, it can be cast as
	$$
		\rho=B  r^{2 \alpha} \int_0^{\sqrt{-2\phi}} \phi^{\beta}\left(-\dfrac{\nu^2}{2\phi}-1\right)^{\beta} \nu^{2\alpha +2 } {\mathrm{d}}\nu,
	$$
	where, by making the substitution $x={\nu^2}/{\phi}$, the integral becomes
	$$
		\rho=B  r^{2 \alpha} \phi^{\beta + \alpha + 1} 2^{\alpha +1 }
		\sqrt{\dfrac{\phi}{2}} \int_0^{-1} {(-x-1)}^{\beta} x^{\alpha +1/2 } {\mathrm{d}}x,
	$$
	Again, the last integral is a constant. With this in mind and organizing the terms, we have
	\begin{equation}\label{dens}
		\rho=C r^{2 \alpha} \phi^{\beta + \alpha + 3/2}
	\end{equation}
	Now it is possible to calculate the potential through the Poisson equation,
	$$
		{\nabla}^2 \phi = 4 \pi G \rho= 4 \pi C  G r^{2 \alpha} \phi^{\beta + \alpha + 3/2}
	$$	
	Since $\alpha$ and $\beta$ are parameters, it is straightforward  to prove that
	$$
		\phi =- \dfrac{\phi_o}{1+r/a},
	$$
	is a solution of the equation for $\alpha=-1/2$, $\beta=2$ and $ 4 \pi C  G = -{2}/{{\phi_o}^{2}a} $,
	where $a$ is the characteristic radius of the system.

	Now returning to the expression (\ref{dens}) of
	the density and thus using the values $\alpha=-1/2$ and $\beta=2$, we can compute the constant $A$:
	$$
		C=2\pi A  \int_0^{\pi}{\mathrm{d}}\eta \int_0^{-1} {(-x-1)}^{2} {\mathrm{d}}x=-\dfrac{2 \pi^2 A}{3},
	$$
	then
	$$
		C=-\dfrac{2 \pi^2 A}{3}=-\dfrac{1}{2\pi G {\phi_o}^{2} a},
	$$
	which lead us to
	\begin{equation}
		A=\dfrac{3}{4 \pi^3 {\phi_o}^{2} a G }
	\end{equation}

	In summary, we can establish  that the distribution function, the gravitational potential and the mass density for the Hernquist model are given by
	\begin{equation}{\label{cdf}}
		\DF(\varepsilon,L)=\dfrac{3}{4 \pi^3 {\phi_o}^{2} a G }{\varepsilon}^{2} L ^ {-1}
	\end{equation}
	\begin{equation}
		\phi =- \dfrac{\phi_o}{1+r/a}
	\end{equation}
	\begin{equation}
		\rho=-\dfrac{1}{2\pi G {\phi_o}^{2} a} r^{-1} {\phi}^{3}=\dfrac{\phi_o}{2\pi G a r} \left( \dfrac{1}{1+r/a}\right)^{3}
	\end{equation}

\section{Demonstration of  Lemma \ref{Lema1}}\label{appendix2}

In this appendix we provide a proof by reductio ad absurdum of lemma \ref{Lema1}, used to obtain theorem \ref{Teorema1}:
\begin{lema}\label{Lema1}
		If $\DF$ is a DF satisfying the self-gravitation equations (\ref{tt-2}), (\ref{tr-2}) and (\ref{tte-2}),
then $\lim\limits_{L\rightarrow 0}\left(L^2\DF\right)=0$.
\end{lema}

\textit{Proof}. If one supposes that
$$
\lim\limits_{L\rightarrow 0}\left(L^2\DF\right)\neq0,
$$
then, from the definition of limit, for every $\delta>0$ there exists
$\epsilon>0$ and $L_0$ such that $0<L_{0}<\delta$ and $\DF(E,L_{0})>\epsilon{L_{0}}^{-2}$.

On the other hand, since $\DF$ must be a continuous function, then $L^2\DF$ is
a continuous function too, so there exists a
region centered in $L_{0}$ such that $\DF>\epsilon{L}^{-2}$, i.e.
$\DF>\epsilon{L}^{-2}$ for every $L$ belonging to $L_{0}-\delta L<L<L_{0}+\delta L$.

All of the above holds
for every choice of $0<\delta<\delta L$. Then, if we choose
$\delta$ in such a way that $0<L<\delta$ and, therefore, $L$ falls inside the interval $(L_{0}-\delta L,L_{0}+\delta L)$, then
for such $\delta$ there exists an $\epsilon>0$ such that whenever $0<L<\delta$ we have $\DF>\epsilon L^{-2}$.

Now, by choosing  $L_m$ to be smaller than $\delta$, we can write
$$
\int^{L_m}_{0}
\frac{\DF(E,L) L\mathrm{d} L}{\sqrt{{L_m}^2-L^2}}\geq \epsilon\int^{L_m}_{0}
\frac{\mathrm{d} L}{L\sqrt{{L_m}^2-L^2}}.
$$
Note that the right hand side integral does not converge and the left hand side integral must
converge since $\rho$, given by (\ref{rho1}), is finite. This means that the relation above is an absurd,
which leads us to state that
\begin{multline*}
\hfill\lim\limits_{L\rightarrow 0}\left(L^2\DF\right)=0. \hfill \blacksquare
\end{multline*}

\footnotesize


\end{document}